\documentclass[12pt]{article} 

\usepackage{latexsym} 	
\usepackage{amssymb}  	
\usepackage{amsmath}  	
\usepackage{amsbsy}
\usepackage{epsfig}     
\usepackage{a4}     

\allowdisplaybreaks

\newcommand{\om}{\omega}

\newcommand{\tr}{\mbox{tr}}

\newcommand{\id}{1\!\!1}

\newcommand{\cE}{{\cal E}}
\newcommand{\cK}{{\cal K}}
\newcommand{\cM}{{\cal M}}
\newcommand{\bra}{\langle}
\newcommand{\ket}{\rangle}

\newcommand{\half}{\frac{1}{2}}

\newcommand{\vecnul}{{\mathbf 0}}

\newcommand{\kv}{{\mathbf k}}  
\newcommand{\pv}{{\mathbf p}}

\newcommand{\rv}{{\mathbf r}}
\newcommand{\xv}{{\mathbf x}}

\newcommand{\Kslash}{\,/\!\!\!\!\!\,K}

\newcommand{\be}{\begin{equation}}
\newcommand{\ee}{\end{equation}}
\newcommand{\bea}{\begin{eqnarray}}
\newcommand{\eea}{\end{eqnarray}}
\newcommand{\bean}{\begin{eqnarray*}}
\newcommand{\eean}{\end{eqnarray*}}
\newcommand{\nn}{\nonumber}
\newcommand{\hm}{\hspace*{-0.6cm}}

\newcommand{\gm}{\gamma}

\renewcommand{\theequation}{\arabic{section}.\arabic{equation}}

\begin{document}

\title{
\vskip -120pt
{\begin{normalsize}
\mbox{} \hfill SWAT 06/483\\
\mbox{} \hfill hep-lat/0612007 \\
\vskip  100pt
\end{normalsize}}
{\bf\Large
Meson spectral functions with chirally symmetric lattice fermions 
}
\author{
\addtocounter{footnote}{2}
Gert Aarts\thanks{email: g.aarts@swan.ac.uk}
 {} and
Justin Foley\thanks{email: j.foley@swansea.ac.uk}
 \\ {} \\
{\normalsize (UKQCD Collaboration)}
 \\ {} \\
{\em\normalsize Department of Physics, Swansea University}
\\
{\em\normalsize Singleton Park, Swansea, SA2 8PP, United Kingdom}
}
}
\date{December 7, 2006}
\maketitle
\begin{abstract}

In order to enhance our understanding of spectral functions in lattice 
QCD obtained with the help of the Maximum Entropy Method, we study meson 
spectral functions for lattice fermions with chiral symmetry. In 
particular we analyse lattice artefacts for standard overlap, 
overlap hypercube and domain wall fermions in the free field limit.
 We also present first results for pseudoscalar spectral 
functions in dynamical QCD with $2+1$ flavours of domain wall fermions, 
using data generated by the UKQCD and RBC collaborations on QCDOC 
machines.

\end{abstract}

\newpage


                                                                     
\section{Introduction}
\setcounter{equation}{0}

At zero temperature, the spectrum of QCD is encoded in hadronic spectral 
functions. Groundstates, excited states, decay widths and continuum 
contributions can, in principle, be extracted from spectral functions in 
different channels. Similarly, at finite temperature and/or density, 
medium modification of hadrons, the rate of photon and dilepton 
production, and hydrodynamical response functions can be obtained from the 
appropriate spectral functions.

Spectral functions are inherently real-time correlation functions and
therefore difficult to obtain using standard lattice QCD data analysis
techniques. Since euclidean lattice correlation functions are determined
numerically on a finite number of points in imaginary time only, the
analytical continuation to real time is classified as an ill-posed
problem. In the past few years significant progress in the extraction of
spectral functions from lattice QCD has come from the application of the
Maximum Entropy Method (MEM) to this problem. MEM has been applied in many
branches of science (see e.g.\ Ref.\ \cite{Jarrell} for a review), a
thorough review focussing on lattice QCD can be found in Ref.\
\cite{Asakawa:2000tr}.

In order to interpret hadronic spectral functions obtained in lattice QCD, 
it is important to understand how lattice artefacts will appear. Lattice 
artefacts are expected to be present at large frequencies and may be 
studied in perturbation theory. Free lattice meson spectral functions have 
been studied at zero momentum for Wilson and hypercube fermions 
\cite{Karsch:2003wy} and at nonzero momentum for Wilson and staggered 
fermions \cite{Aarts:2005hg}. In this paper our first goal is to study 
lattice meson spectral functions for chirally symmetric fermions, 
specifically overlap \cite{Narayanan:1992wx,Neuberger:1997fp}, domain wall 
\cite{Kaplan:1992bt,Shamir:1993zy,Furman:1994ky}, and overlap hypercube 
\cite{Bietenholz:1998ut,Bietenholz:1999km} fermions. Furthermore we 
present results for meson spectral functions obtained with the Maximum 
Entropy Method in QCD with $2+1$ flavours of domain wall fermions, using 
data generated by the UKQCD and RBC collaborations on QCDOC machines.

The paper is organized as follows. In the next section we derive general 
expressions for free meson spectral functions on a finite lattice, 
independent of the particular fermion formulation that is used. In Section 
\ref{secov} we specialize to overlap fermions and compare the resulting 
spectral functions, and in particular lattice artefacts, with continuum 
and staggered spectral functions. This analysis is extended to 
domain wall fermions in Section \ref{secdw} and to overlap hypercube 
fermions in Section \ref{secovHF}. In Section \ref{secUKQCD} we present 
first results for pseudoscalar spectral functions in QCD with 
dynamical domain wall fermions. We find good agreement between the MEM 
results and the groundstate mass obtained with conventional cosh fits. We 
also argue that the structure seen at larger energies is consistent with 
lattice artefacts found in the first part of the paper, but a quantitative 
comparison requires further study.

\section{Spectral functions}
\setcounter{equation}{0}

We start with a brief summary of well-known relations 
\cite{Asakawa:2000tr}.
Euclidean meson correlators are defined by
\be
G_H(\tau,\xv) = \bra J_H(\tau,\xv) J_H^\dagger(0,\vecnul)\ket,
\ee
where $J_H(\tau,\xv)=\bar\psi(\tau,\xv)\Gamma_H\psi(\tau,\xv)$ and 
$\Gamma_H=\{\id,\gamma_5, \gamma^\nu, \gamma^\nu\gamma_5 \}$ for 
the scalar, pseudoscalar, vector and axial vector channels respectively.
In general euclidean (and other) correlation functions are related to 
their spectral function via a dispersion relation in momentum space,
\be
G_H(z,\pv) = \int_0^\infty \frac{d\om}{2\pi} 
\frac{\rho_H(\om,\pv)}{\om-z},
\ee
 where $z$ is the frequency extended into the complex plane. Equating $z$ 
to $i\om_n$, where $\om_n=2\pi n T$ ($n\in \mathbb{Z}$) is the 
Matsubara frequency, yields the euclidean correlator at finite 
temperature $T$.
In imaginary time this relation reads
\be
\label{eqGKr}
G_H(\tau,\pv) = \int_0^\infty \frac{d\om}{2\pi}\, 
K(\tau,\om)\rho_H(\om,\pv),
\ee
with the kernel
\be
 K(\tau,\om) =\frac{\cosh[\om(\tau-1/2T)]}{\sinh(\om/2T)}.
\ee
At zero temperature, this kernel reduces to $K(\tau,\om) = 
e^{-\omega\tau}$.

In this section we derive a general expression for meson spectral 
functions on an isotropic lattice with $N_\sigma^3\times N_\tau$ sites, by 
writing the euclidean correlators in the form (\ref{eqGKr}) and 
identifying the lattice spectral function from that expression. 
We use periodic boundary conditions in 
space, $k_i = 2\pi n_i/N_\sigma$ with $n_i = -N_\sigma/2+1, -N_\sigma/2+2, 
\ldots, N_\sigma/2-1, N_\sigma/2$ for $i=1, 2, 3$, and antiperiodic 
boundary conditions in imaginary time, $k_4 = \pi(2n_4+1)/N_\tau$ with 
$n_4 = -N_\tau/2+1, -N_\tau/2+2, \ldots, N_\tau/2-1, N_\tau/2$. Lattice 
units $a=1$ are used throughout.

The correlators we are interested in are of the form
\be
G_H(\tau,\pv) = -\frac{N_c}{N_\sigma^3} \sum_\kv \tr\, S(\tau,\kv) 
\Gamma_H S(-\tau,\pv+\kv) \Gamma_H,
\label{eqG0}
\ee
where $S(\tau,\kv)$ is the fermion propagator and $N_c$ denotes the number 
of colours.
 In order to derive compact expressions for lattice meson spectral 
functions, starting from Eq.\ (\ref{eqG0}), we first discuss a 
generic fermion propagator, without specifying a particular 
formulation. We consider the following fermion propagator
\be
 S(K) = \frac{1}{D(K)} \left[ -i\sum_{\nu=1}^4 C_\nu(K) \gamma_\nu \sin 
k_\nu +  m(K) \right],
\ee
where $K$ denotes the four-momentum.
 The functions $C_\nu(K)$, $m(K)$ and $D(K)$ depend on the fermion 
formulation, but they are all invariant under $k_\nu\to -k_\nu$. In order 
to be able to use Eq.\ (\ref{eqG0}), we construct the fermion propagator 
in the mixed representation,
 \be
 S(\tau,\kv) = \frac{1}{N_\tau}\sum_{k_4} e^{ik_4\tau} S(K).
\ee
We assume that $S(K)$ has a single pole at $k_4=\pm iE_\kv$, determined by
$D(iE_\kv,\kv)=0$. In the case of more than one pole, a summation over 
the poles is required. This yields \cite{Carpenter:1984dd}
\be 
 \label{eqSCB}
 S(\tau, \kv) = \gamma_4 S_4(\tau, \kv) +
 \sum_{i=1}^3 \gamma_i S_i(\tau, \kv) +\id S_u(\tau, \kv),
\ee
where
\bea
\nn
S_4(\tau, \kv) =&&\hm S_4(\kv)\cosh(\tilde\tau E_\kv), \\
\nn
S_i(\tau, \kv) =&&\hm S_i(\kv)\sinh(\tilde\tau E_\kv), \\
S_u(\tau, \kv) =&&\hm S_u(\kv)\sinh(\tilde\tau E_\kv).
\eea
Here $0\leq \tau<N_\tau=1/T$ and $\tilde\tau 
= \tau-1/2T$. The momentum-dependent coefficients read
\bea
\nn
S_4(\kv) =&&\hm \frac{C_4(iE_\kv,\kv)}{2\cE_\kv}
\frac{\sinh E_\kv}{\cosh(E_\kv/2T)},
\\\nn
S_i(\kv) =&&\hm \frac{C_i(iE_\kv,\kv) }{2\cE_\kv}
\frac{i\sin k_i}{\cosh(E_\kv/2T)},
\\
S_u(\kv) =&&\hm - \frac{m(iE_\kv,\kv)}{2\cE_\kv\cosh(E_\kv/2T)},
\label{eqsss}
\eea
where
\be
\label{eqres}
 \frac{1}{2\cE_\kv} = 
 i \mathop{\mbox{Res}}_{k_4=iE_\kv} \frac{1}{D(K)}.
\ee
 The propagator satisfies $ S(-\tau,\kv) = \gm_5S^\dagger(\tau,\kv) 
\gm_5$.

For reference, we note that for fermions in the continuum one finds the 
same propagator, with the replacements $\cE_\kv\to E_\kv$, $\sinh E_\kv\to 
E_\kv$, $\sin k_i\to k_i$, $m(iE_\kv,\kv)\to m$ and $C_\nu\to 1$.

Inserting Eq.\ (\ref{eqSCB}) in Eq.\ (\ref{eqG0}) gives the euclidean 
correlator
\bea
\nn
\!\!\!\!
G_H(\tau,\pv) =  \frac{4N_c}{N_\sigma^3} \sum_\kv \Big[
&&\hm
a_H^{(1)} S_4(\tau,\kv)S_4^\dagger(\tau,\rv)
\\ &&\hm
- a_H^{(2)} \sum_i S_i(\tau,\kv)S_i^\dagger(\tau,\rv)
- a_H^{(3)} S_u(\tau,\kv)S_u^\dagger(\tau,\rv) \Big], 
\label{eqG}
\eea
where $\rv=\pv+\kv$. The coefficients $a_H^{(i)}$ are given in Table 
\ref{table1}.\footnote{We use Minkowski gamma-matrices to label the 
channels, see Ref.\ \cite{Aarts:2005hg} for further details.}

\begin{table}[t]
\begin{center}
\begin{tabular}{|c|c|c|r|r||c|c|c|r|r|}
\hline
& $\Gamma_H$    & $a_H^{(1)}$ & $a_H^{(2)}$ & $a_H^{(3)}$ & 
& $\Gamma_H$    & $a_H^{(1)}$ & $a_H^{(2)}$ & $a_H^{(3)}$ \\
\hline
$\rho_{\rm S}$  &$\id$          & $1$   & $-1$  & $1$  & 
$\rho_{\rm PS}$ & $\gamma_5$    & $1$   & $-1$  & $-1$  \\
$\rho^{00}$     &$\gamma^0$     & $1$   & $1$   & $1$   &
$\rho^{00}_5$   &$\gamma^0\gamma_5$     & $1$   & $1$   & $-1$  \\
$\rho^{ii}$     &$\gamma^i$     & $3$   & $-1$  & $-3$  &
$\rho^{ii}_5$   &$\gamma^i\gamma_5$     & $3$   & $-1$  & $3$   \\
$\rho_{\rm V}$  &$\gamma^\mu$   & $2$   & $-2$  & $-4$  &
$\rho_{\rm A}$  &$\gamma^\mu\gamma_5$   & $2$   & $-2$  & $4$   \\
\hline
\end{tabular}
 \caption{Coefficients $a_H^{(i)}$ for free spectral functions in
 different channels $H$. In the case of $\gamma^i$ and $\gamma^i\gamma_5$,
the sum is taken over $i=1,2,3$; $\rho_{\rm V} = \rho^{ii} 
-\rho^{00}$ and $\rho_{\rm A} = \rho^{ii}_5 -\rho^{00}_5$.
}
\label{table1}
\end{center}
\end{table}

We will now extract the lattice meson spectral functions. It is 
straightforward to write the above expression for $G_H(\tau,\pv)$ as
\be
G_H(\tau,\pv) =  \int_{0}^\infty \frac{d\om}{2\pi}\,
K(\tau,\om) \rho^{\rm lattice}_{H}(\om,\pv),
\ee
and identify the expressions for the lattice spectral functions 
\cite{Aarts:2005hg}
\bea
&&\hm
\rho^{\rm lattice}_H(\om,\pv) =
\frac{4\pi N_c}{N_\sigma^3}\sum_\kv \sinh\left(\frac{\om}{2T}\right)
\bigg\{
\nn \\ &&\hm \;\;\;\;\;\;\;\;
\bigg[
a_H^{(1)} S_4(\kv)S^\dagger_4(\rv)
+ a_H^{(2)} \sum_i S_i(\kv)S^\dagger_i(\rv)
+ a_H^{(3)} S_u(\kv)S^\dagger_u(\rv) \bigg]
\delta(\om +E_\kv-E_\rv)
\nn \\&&\hm \;\;\;\;\;\;\;\;
+ \bigg[
a_H^{(1)} S_4(\kv)S^\dagger_4(\rv)
- a_H^{(2)} \sum_i S_i(\kv)S^\dagger_i(\rv)
- a_H^{(3)} S_u(\kv)S^\dagger_u(\rv) \bigg]
\delta(\om -E_\kv-E_\rv)
\nn \\ &&\hm \;\;\;\;\;\;\;\;
+ (\om\to-\om)
\bigg\}.
\label{eqrhoW}
\eea
This expression is the immediate counterpart of the continuum result 
\cite{Aarts:2005hg}
\bea
&&\hm
\rho^{\rm cont}_H(\om,\pv) =
2\pi N_c \int \frac{d^3k}{(2\pi)^3} 
\frac{1}{E_\kv E_\rv}
\bigg\{
\nn \\ &&\hm \;\;\;\;\;\;\;\;
\left[ n_F(E_\kv)-n_F(E_\rv) \right]
\left[ a_H^{(1)} E_\kv E_\rv + a_H^{(2)} \kv\cdot\rv
+ a_H^{(3)} m^2 \right]  \delta(\om +E_\kv-E_\rv)
\nn \\&&\hm \;\;\;\;\;\;\;\;
+ 
\left[ 1- n_F(E_\kv) - n_F(E_\rv) \right]
\left[ a_H^{(1)} E_\kv E_\rv - a_H^{(2)} \kv\cdot\rv - a_H^{(3)} 
m^2\right] \delta(\om -E_\kv-E_\rv)
\nn \\ &&\hm \;\;\;\;\;\;\;\;
- (\om\to-\om)
\bigg\},
\eea
where $n_F(\om)=1/[\exp(\om/T)+1]$ is the Fermi distribution, as can be 
seen by making the appropriate substitutions. At zero temperature, only 
the ``1'' in the second term survives.  
 In the continuum the remaining three-dimensional integral can be carried 
out for arbitrary external momentum and quark mass. Analytical expressions 
for continuum meson spectral functions can be found in Ref.\ 
\cite{Aarts:2005hg}. Since the expressions are lengthy, we give here 
the result at vanishing external momentum only,
 \bea
 \rho_H^{\rm cont} (\om,\vecnul) = &&\hm 
 \Theta(\om^2-4m^2)\frac{N_c}{8\pi}
 \sqrt{1-\frac{4m^2}{\om^2}}\left[1-2n_F\left(\frac{\om}{2}\right)\right]
\nn \\
&&\hm 
 \left[ \om^2\left( a_H^{(1)} -a_H^{(2)} \right)
+4m^2\left( a_H^{(2)} -a_H^{(3)} \right) \right]
\nn \\
&&\hm 
-4\pi \om\delta(\om) N_c
\int \frac{d^3k}{(2\pi)^3}\, \frac{ n'_F(\om_\kv)}{\om_\kv^2}
\left[ 
k^2\left(a_H^{(1)} + a_H^{(2)}\right) +
m^2 \left( a_H^{(1)} + a_H^{(3)}\right) 
\right].
\nn \\
\label{eqcont}
\eea
 The first term contributes above threshold ($\om>2m$), while the 
contribution proportional to $\om\delta(\om)$ is related to conserved 
quantities (see e.g.\ Ref.\ \cite{Aarts:2002cc} in relation to 
transport coefficients). Note that spectral functions are odd,  
$\rho_H(-\om,\pv)=-\rho_H(\om,\pv)$.

On the lattice the spectral functions (\ref{eqrhoW}) can in general not be 
evaluated analytically. Instead, in the following sections we give the 
explicit expressions for the free fermion dispersion relation $E_\kv$, the 
coefficients $S_4(\kv)$, $S_i(\kv)$ and $S_u(\kv)$, and the residue 
$\cE_\kv$, for overlap, domain wall and overlap hypercube fermions. We use 
those in Eq.\ (\ref{eqrhoW}) to compute lattice spectral functions by 
performing the spatial lattice sum over $\kv$ numerically, using the same 
approach as in Refs.\ \cite{Karsch:2003wy,Aarts:2005hg}.

\section{Overlap fermions}
\setcounter{equation}{0}
\label{secov}

The massless overlap (Neuberger) operator is given by
\be
D_{N} = \mu \left( 1+ \frac{X}{\sqrt{X^\dagger X}}\right), 
\;\;\;\;\;\;\;\;
X = D_W-\mu,
\ee
 where $D_W$ the usual Wilson-Dirac operator and $\mu$ is a 
constant.\footnote{In other studies the coefficient $\mu$ is sometimes 
denoted by $\rho$.} In momentum space, $X$ reads
\be
X(K) =  i\sum_\nu \gamma_\nu \sin k_\nu + b(K),
\;\;\;\;\;\;\;\;
b(K) = r\sum_\nu\left(1-\cos k_\nu\right) - \mu.
\ee
The corresponding fermion propagator is
\be
S(K) = \frac{1}{2\mu}\left( 
\frac{-i\sum_\nu \gamma_\nu \sin k_\nu}{\om(K) +b(K)} + 1\right),
\ee
with
\be
\om(K) = \sqrt{X^\dagger X} = \sqrt{\sum_\nu \sin^2 k_\nu + b^2(K)}.
\ee
 The poles of the propagator are determined by $\om(K) + 
b(K) = 0$, which yields
\be
\sum_\nu \sin^2 k_\nu = 0, \;\;\;\;\;\;\;\; \;\;\;\; \;\;\;\; b(K)<0.
\ee
Writing $k_4=iE_\kv$ gives the dispersion relation
 \be
\cosh E_\kv = \sqrt{1+\cK_\kv^2}, 
\;\;\;\;\;\;\;\;\;\;\;\;\;\;\;\;
b(iE_\kv, \kv)<0,
\ee
where we defined
\be
\label{eqmbar}
\cK_\kv^2 = \sum_i \sin^2 k_i.
\ee
 The constraint $b<0$ arises from the square root in the definition of 
$\om(K)$.\footnote{Provided that $0<\mu<2r$, there is no pole at 
$k_4=\pi-iE_\kv$, since then the constraint cannot be met.} Due to this 
constraint, there is only a pole when $\kv$ is not too large, and there 
are no solutions near the edges of the Brillouin zone. Apart from this, 
the dispersion relation is identical to the one for naive fermions.

Meson spectral functions for free massless overlap fermions take the 
form (\ref{eqrhoW}) of the previous section, with the now explicitly 
determined functions (\ref{eqsss})
\be
 C_\nu(iE_\kv,\kv) = 1,
\;\;\;\; \;\;\;\; \;\;\;\; \;\;\;\;
 S_u(\kv)=0,
\ee
and the residue (\ref{eqres})
\be
\frac{1}{\cE_\kv} = \frac{\om(iE_\kv,\kv)}{\mu \sinh E_\kv \cosh E_\kv}.
\ee
 Chiral symmetry is manifest in the meson spectral functions, since the 
expressions in the scalar (vector) and the pseudoscalar (axial vector) 
channel only differ with respect to the coefficient $a_H^{(3)}$, see Table 
\ref{table1}. Since $S_u=0$, dependence on $a_H^{(3)}$ vanishes for 
massless overlap fermions.

The extension to massive overlap fermions is straightforward. 
To include the mass, the overlap operator is changed to
\be
 D_{{\rm ov},m_0} = \left(1-\frac{m_0}{2\mu}\right) D_{N} + m_0,
\ee
and the corresponding propagator is
\be
\label{eqgen}· 
S(K) = 
\half \frac{(\mu-m_0/2)\left[ -i\sum_\nu \gamma_\nu \sin k_\nu 
+ b(K) \right] + (\mu+m_0/2) \om(K) }{ (\mu^2+m_0^2/4)\om(K)  + 
 (\mu^2-m_0^2/4) b(K)}.
\ee
We find that the pole is determined by 
\be
\label{eqpole}· 
\sum_\nu \sin^2 k_\nu = -\bar m_0^2 b^2(K), 
\;\;\;\;\;\;\;\;\;\;\;\;\; 
\left(\mu^2-m_0^2/4\right) b(K)<0,
\ee
where we defined
\be
\bar m_0 = \frac{\mu m_0}{\mu^2+m_0^2/4}.
\ee
We always consider $m_0$ to be small (in lattice units), while 
$\mu\sim 1$, so that the constraint still implies $b<0$.
Writing again $k_4=iE_\kv$ and solving the resulting quadratic equation 
gives the allowed dispersion relation
\be
 \cosh E_\kv = \frac{1}{1-r^2\bar m_0^2} \left[ -r\bar m_0^2 \cM_\kv +
\sqrt{ (1+\cK_\kv^2)(1-r^2\bar m_0^2) + \bar m_0^2 \cM_\kv^2}\right],
\label{eqEov}
\ee
provided that
\be
 b(iE_\kv, \kv) = -r\cosh E_\kv + \cM_\kv < 0.
\ee
Here we have defined
\be
\label{eqMk}
 \cM_\kv = r - \mu + r\sum_i(1-\cos k_i).
\ee
At zero momentum, the rest mass is determined by
\be
 \cosh E_\vecnul = \frac{1}{1-r^2\bar m_0^2} \left[ r(\mu-r)\bar m_0^2 
+
\sqrt{ 1+ \bar m_0^2\mu(\mu-2r)}\right],
\ee
while at small $\kv$ and $m_0$ Eq.\ (\ref{eqEov}) reduces to
 \be
 \cosh E_\kv = 1+ \frac{1}{2}\left(\kv^2+ m_0^2\right)+ \ldots,
\ee
as expected.

\begin{figure}[!p]
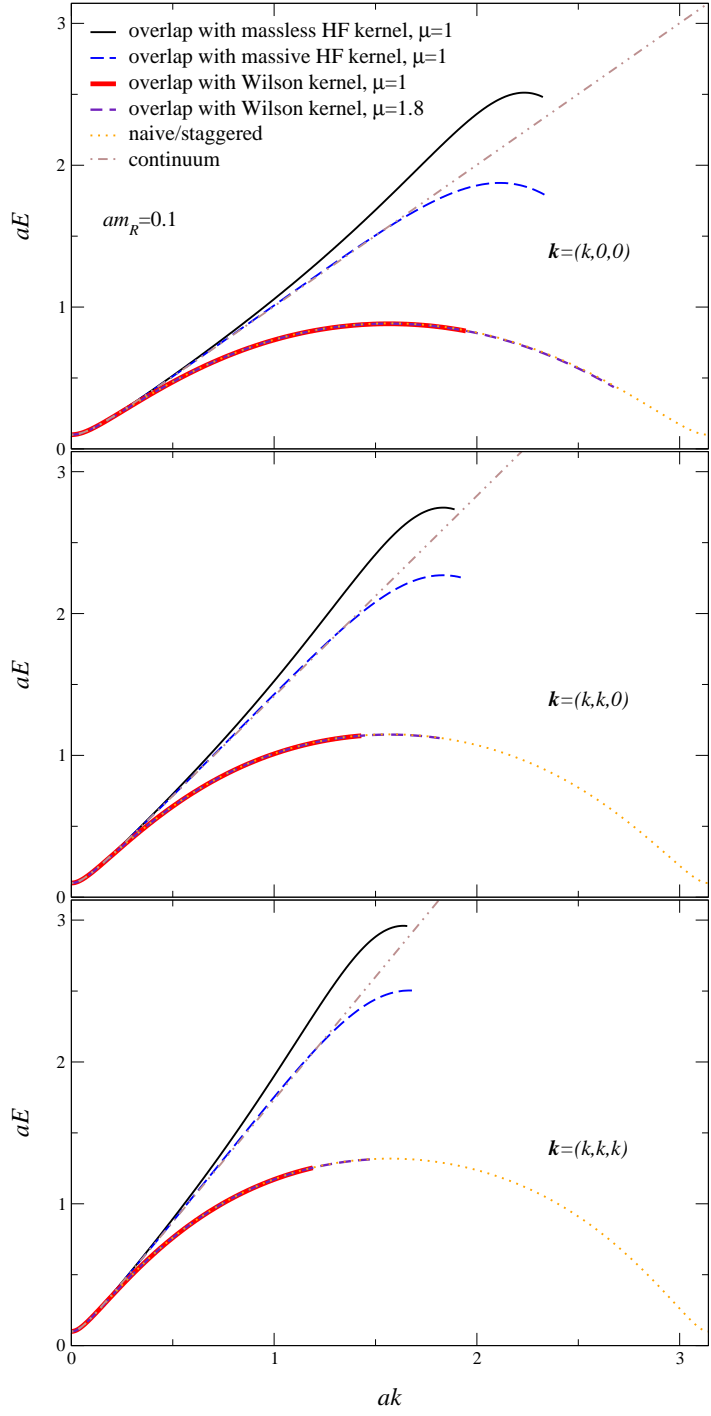

\begin{center}
\epsfig{figure=disp100_2.eps,width=9.3cm}\\
\vspace*{-0.13cm}
\epsfig{figure=disp110_2.eps,width=9.3cm}\\
\vspace*{-0.13cm}
\epsfig{figure=disp111_2.eps,width=9.3cm}
\end{center}
\vspace*{-0.3cm}
 \caption{Dispersion relation along three directions in the Brillouin zone 
for massive overlap fermions with rest mass $m_R=0.1$ using a 
massive/massless HF kernel with $\mu=1$ and a standard Wilson kernel with 
$\mu=1, 1.8$. For comparison the naive and continuum dispersion relation 
are shown as well.
  }
\label{figdisp}
\end{figure}

We compare the massive overlap dispersion relation with the 
naive (or staggered) dispersion relation,
\be
\cosh E_\kv = \sqrt{1+\cK^2_\kv +m_0^2},
\ee
 and the continuum expression $E_\kv=\sqrt{\kv^2+m_0^2}$ in Fig. 
\ref{figdisp} for two values of the overlap parameter $\mu$ and a rest 
mass $m_R\equiv E_\vecnul=0.1$ (the HF kernel shown as well will be 
discussed below). Throughout this paper we take $r=1$. We find that the 
overlap dispersion relation can hardly be distinguished from the naive one 
for the value of $m_R$ shown here. In the overlap case the dispersion 
relation terminates before the edge of the Brillouin zone due to the 
constraint $b<0$. The momentum value of the endpoint depends on the 
overlap parameter $\mu$ and increases with increasing $\mu$.

\begin{figure}[t]
\centerline{\epsfig{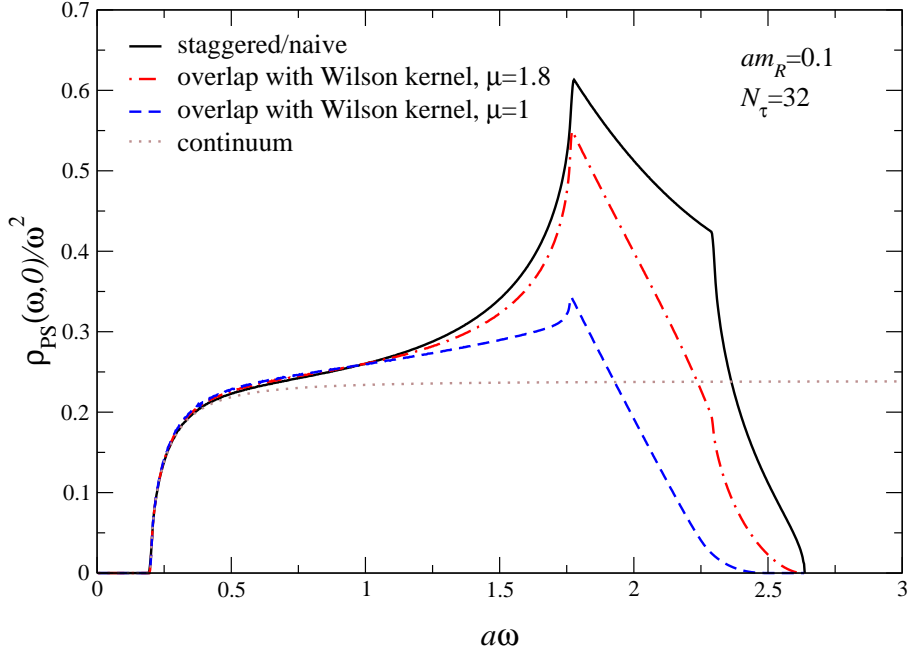}}
\caption{Pseudoscalar spectral functions $\rho_{\rm 
PS}(\om,\vecnul)/\om^2$ for staggered and standard overlap 
fermions with $\mu=1$, $1.8$, and $m_R=0.1, N_\tau=32$.
}
\label{figrho_PS_stag_OV_1}
\end{figure}

The coefficients in the meson spectral functions now read
\be
\label{eqCm}
 C_\nu(iE_\kv,\kv) = 1-\frac{m_0}{2\mu}, 
 \;\;\;\; \;\;\;\;
 m(iE_\kv,\kv) = \frac{m_0}{\mu+m_0/2} \om(iE_\kv,\kv),
\ee
and the residue is
\be
\frac{1}{\cE_\kv} = 
 \frac{\mu}{\mu^2+m_0^2/4}
\frac{1}{\cosh E_\kv + \bar m_0^2 r b(iE_\kv,\kv)}
\frac{\om(iE_\kv,\kv)}{\sinh E_\kv}.
\ee
Comparison between these functions at small momentum 
and their continuum counterparts shows that there is a multiplicative 
renormalization of the fermion propagator at finite $m_0$. This is not 
unexpected since the kinetic term in (\ref{eqgen}) has a nonstandard 
normalization. One way to write the renormalization factor is to compare 
the expressions at zero spatial momentum. Explicitly, the renormalization 
factor is given by 
\be
C_\nu(iE_\vecnul,\vecnul) \frac{E_\vecnul}{\cE_\vecnul}.
\ee
 At small $K$ and $m_0$, the euclidean fermion propagator (\ref{eqgen}) is 
approximately given by
 \be
S(K) \approx \frac{\mu^3(\mu-m_0/2)}{\left(\mu^2 + m_0^2/4\right)^2} 
\left[ \frac{-i\Kslash + m_0}{K^2+m_0^2} + \frac{1}{2\mu} \right].
\ee
 The multiplicative prefactor goes to 1 in both the chiral and the 
continuum limit.

We now have all the ingredients to compute spectral functions with free 
overlap fermions. In Fig.\ \ref{figrho_PS_stag_OV_1} we show the 
pseudoscalar spectral function at zero momentum for overlap fermions with 
$\mu=1$ and $1.8$, staggered fermions and the continuum result 
(\ref{eqcont}). All spectral functions increase in the same manner beyond 
the threshold $\om=2m_R$, provided that $m_R$ is small. At larger 
frequencies, effects due to the deviation of the continuum and lattice 
dispersion relation become visible. The first cusp is due to the maximal 
lattice energy reached at $\kv=(\pi/2,0,0)$ (cf.\ Fig.\ \ref{figdisp} 
top), which yields a cusp at $a\om \approx 2 \cosh^{-1} \sqrt{2} \approx 
1.76$. The difference in height of the spectral functions is due to 
the different residues, which depends on $\mu$ for 
overlap fermions and for staggered fermions is given by 
\cite{Aarts:2005hg}
 \be
\frac{1}{\cE_\kv} = \frac{1}{\cosh E_\kv\sinh E_\kv}.
\ee
 For staggered fermions, there is a second cusp due to the maximal lattice 
energy reached at $\kv=(\pi/2,\pi/2,0)$. For overlap fermions this cusp is 
absent (for $\mu=1$) or less pronounced (for $\mu=1.8$), since the 
constraint has terminated the dispersion relation (cf.\ Fig.\ 
\ref{figdisp} middle). The maximal energy is given by $\om_{\rm 
max}=2E_{\kv\rm max}$, which for staggered fermions is reached at 
$\kv=(\pi/2,\pi/2,\pi/2)$ and given by $a\om \approx 2 \cosh^{-1}2 \approx 
2.63$. For overlap fermions, this region is again modified due to the 
constraint.

\begin{figure}[t]
\centerline{\epsfig{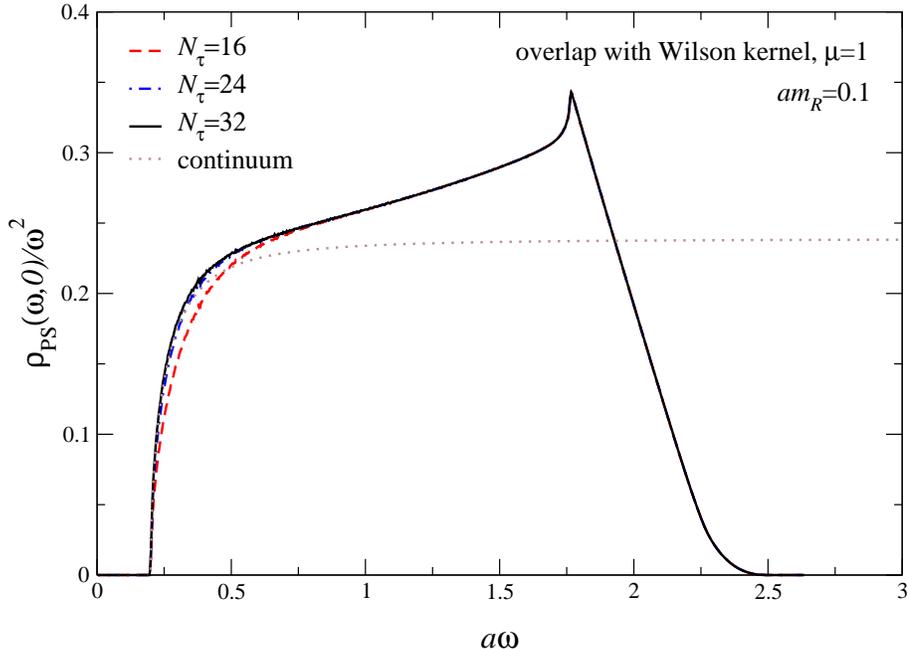}}
\caption{$N_\tau$ dependence of the pseudoscalar spectral function $\rho_{\rm 
PS}(\om,\vecnul)/\om^2$ for 
standard overlap fermions ($\mu=1$, $m_R=0.1$). 
}
\label{figrho_PS_Nt_OV_1}
\end{figure}

The effect of finite $N_\tau$ is shown in Fig.\ \ref{figrho_PS_Nt_OV_1} 
for overlap fermions with $\mu=1$ and is seen to be negligible for large 
enough values of $N_\tau$.

Finally, lattice spectral functions at nonzero momentum are similar to the 
continuum ones in the small frequency region $a\om\lesssim 0.5$ and 
lattice artefacts at larger $\om$ are not affected by the external 
momentum, provided that the external momentum is small (see Ref.\ 
\cite{Aarts:2005hg} for further details).

\section{Domain wall fermions}
\setcounter{equation}{0}
\label{secdw}

The four-dimensional massive domain wall propagator, assuming an infinite 
extent of the fifth dimension, reads
 \be 
 S(K) = \frac{-i\sum_\nu \gamma_\nu \sin k_\nu + 
 m_0 \left(1-|W|e^{-\alpha}\right) }{e^{\alpha}|W|-1 + 
m_0^2\left(1-|W|e^{-\alpha}\right) }, 
\ee 
 where 
\bea 
\cosh \alpha = &&\hm \frac{1+W^2 + \sum_\nu \sin^2 k_\nu }{2|W|}, \\ 
W = &&\hm 1-\mu + \sum_\nu(1-\cos k_\nu). 
\eea
The domain wall height is denoted with $\mu$ and is taken between 0 and 
2. For small momentum and $1<\mu<2$, $W$ is negative (see e.g.\ Ref.\ 
\cite{Capitani:2002mp} for a recent review).

 The dispersion relation is determined by the pole in the propagator at 
$k_4=iE_\kv$.  Again we find a quadratic equation for $\cosh E_\kv$, with 
the allowed solution
\be
\label{eqEdw}
\cosh E_\kv = \frac{x_\kv +  (1+m_0^2)\sqrt{y_\kv}}{z_\kv},
\ee
with
\bea
\nn
x_\kv = &&\hm -2m_0^2\left(1+\cM_\kv\right)\left[ 
1+\cK_\kv^2+\cM_\kv(2+\cM_\kv)\right],\\
\nn
y_\kv = &&\hm \left(1+\cK_\kv^2\right) \left(1-m_0^2\right)^2
+ m_0^2 \left[
1+\cK_\kv^2- \cM_\kv(2+\cM_\kv) \right]^2, 
\\
z_\kv = &&\hm (1-m_0^2)^2 - 4m_0^2 \cM_\kv (2+\cM_\kv),
\eea
where $\cM_\kv$ is given in Eq.\ (\ref{eqMk}) with $r=1$.

For massless domain wall fermions, Eq.\ (\ref{eqEdw}) reduces to
\be
\cosh E_\kv = \sqrt{1+\cK_\kv^2},
\ee
as in the overlap formalism. The rest mass is determined by
\be
\label{eqEdpole}
\cosh E_\vecnul = 
\frac{ 2m_0^2(\mu-2)^3 + 
(1+m_0^2)\sqrt{1+m_0^2(2+\mu(\mu-4)(\mu-2)^2)+m_0^4} }{
1-2m_0^2(2\mu^2-8\mu+7) +m_0^4}.
\ee
Expanding the massive case for small $\kv$ and $m_0$, we find
\be
\cosh E_\kv = 1 + \frac{1}{2}\left( \kv^2+ m_{\rm eff}^2 \right) + 
\ldots,
\ee
where $m_{\rm eff} = (1-w_0^2)m_0$ is the multiplicatively 
renormalized fermion mass. The multiplicative factor reads
\be 
1-w_0^2 = \mu (2-\mu), \;\;\;\;\;\;\;\; w_0=W(0)=1-\mu.
\ee
In the limit that $m_0\ll 1$, $m_R=m_{\rm eff}$.
The constraint on the allowed fermion energies arises in this case from 
the behaviour of the propagator in the fifth direction: demanding  
a normalizable solution of the five-dimensional Dirac equation yields the 
constraint 
\be
|W(iE_\kv,\kv)|<1,
\ee
both in the massless and the massive case.\footnote{We note here that 
$W(iE_\kv,\kv)$ is always larger than $-1$, so that this constraint 
coincides with $b(iE_\kv,\kv)<0$ in the overlap formalism, since $W=1+b$.}

The resulting dispersion relation (\ref{eqEdw}) is indistinguishable from 
the dispersion relation for massive overlap fermions for small values of 
$m_0$, shown in Fig.\ \ref{figdisp}. At finite values of $m_0$, deviations 
are more pronounced for larger values of $\mu$. 

\begin{figure}[t]
\centerline{\epsfig{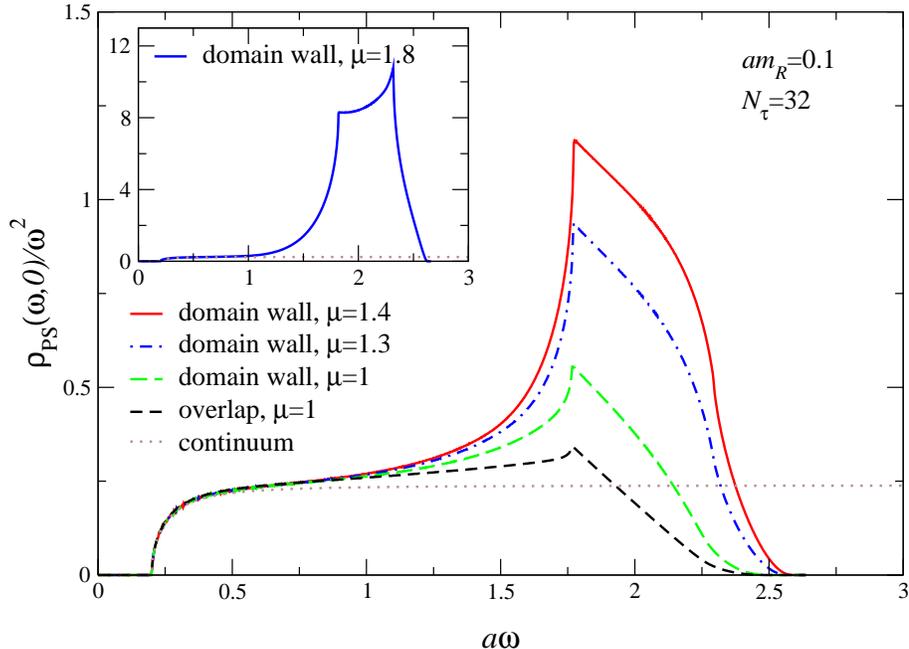}}
\caption{Pseudoscalar spectral functions $\rho_{\rm 
PS}(\om,\vecnul)/\om^2$ for domain wall fermions (with $\mu=1, 1.3, 1.4$ 
and $\mu=1.8$ in the inset) and standard overlap fermions (with $\mu=1$).
}
\label{figrho_PS_DW_OV_1}
\end{figure}

The coefficients in the meson spectral functions read
\be
\label{eqCmdw}
 C_\nu(iE_\kv,\kv) = 1, 
 \;\;\;\; \;\;\;\;
 m(iE_\kv,\kv) = m_0 \left( 1 -\half A_-\right),
\ee
and the residue reads
\be
\frac{1}{\cE_\kv} = 
\frac{2}{A_+\left(1+\cM_\kv\right)-2W 
+ m_0^2\left[A_-\left(1+\cM_\kv\right)-2W\right]} 
\frac{\sqrt{A^2-4W^2}}{\sinh E_\kv}.
\ee
Here we defined
\be
A_\pm = A\pm \sqrt{A^2-4W^2}, 
\;\;\;\;\;\;\;\;
A=1+W^2-\sinh^2 E_\kv+\cK_\kv^2,
\ee
and all quantities are evaluated onshell at $K=(iE_\kv,\kv)$.
In the massless case the residue simplifies considerably to
\be
\frac{1}{\cE_\kv} = \frac{1-W^2(iE_\kv,\kv)}{\cosh E_\kv\sinh E_\kv}.
\ee
 At small $K$ and $m_0$, the euclidean fermion propagator is 
approximately given by
\be
S(K) \approx \left(1-w_0^2\right)
\frac{-i\Kslash + m_{\rm eff} }{ K^2+m_{\rm eff}^2},
\ee
 which suggests a multiplicative renormalization factor $1-w_0^2$. 
However, we have found numerically that corrections to this 
renormalization factor are large for small but finite $m_0$ and larger 
values of $\mu$. From a comparison of the expressions at zero spatial 
momentum, we find the multiplicative factor to be
 \be
 \label{eqmult}
\frac{E_\vecnul}{\cE_\vecnul},
\ee
 which deviates substantially from $1-w_0^2$ for the largest value of 
$\mu$ used here. The resulting spectral functions are shown in Fig.\ 
\ref{figrho_PS_DW_OV_1} for different values of $\mu$. To obtain these 
spectral functions, we fix the rest mass $m_R=0.1$ and solve for the 
bare mass $m_0$ using Eq.\ (\ref{eqEdpole}). Subsequently we find the 
multiplicative factor using (\ref{eqmult}) to properly set the vertical 
scale.

As can be seen in Figs.\ \ref{figrho_PS_DW_OV_1} and 
\ref{figrho_PS_stag_OV_1}, the spectral functions obtained from overlap 
and domain wall fermions differ at larger frequencies. The overlap 
spectral functions show reduced discretization effects. We note that is 
due to the difference in residues, and not because of the dispersion 
relations. Increasing the domain wall height shows that the $\mu$ 
dependence is much stronger than in the overlap case. For $\mu=1.8$ the 
effect is remarkably large.


\section{Overlap hypercube fermions}
\setcounter{equation}{0}
\label{secovHF}

The overlap formalism solves the chirality problem in lattice QCD. 
However, its dispersion relation shows no reduction in discretisation 
effects when compared to e.g.\ staggered fermions. A systematic approach 
to constructing fermion 
actions which have good chiral properties and very small discretisation 
errors is based on the renormalisation group (RG). In this approach one 
aims to approximate so-called `perfect actions', which lie on the 
renormalized trajectory of an RG transformation
\cite{Hasenfratz:1993sp, Hasenfratz:1998bb}. Here we consider a particular 
truncation of a perfect action, the hypercube fermion (HF) action 
introduced in Ref.\ \cite{Bietenholz:1996pf}. This action is obtained from 
an RG transformation which yields an ultralocal action in one 
dimension~\cite{Bietenholz:1995cy}. For a recent review, see 
Ref.~\cite{Bietenholz:2006}.

The hypercube action is written as
\be
 S = \sum_{x,y} \bar\psi(x) \left\{ \sum_\nu\gamma_\nu 
\rho_\nu(x-y)+\lambda(x-y) \right\} \psi(y),
\ee
where the couplings $\rho_\nu$ and $\lambda$ are nonzero only if $x$ and 
$y$ are in the same hypercube. Explicitly, the corresponding Dirac operator 
reads, in momentum space, 
\be
 D_{\rm HF}(K) = i\sum_\nu C_\nu^{\rm HF}(K)\gamma_\nu\sin k_\nu + 
m_{\rm HF}(K),
\ee
with
\be
\label{eqCnu}
 C_\nu^{\rm HF}(K) = 2\rho^{(1)} 
 + 4\rho^{(2)}\sum_{\sigma\neq\nu} \cos k_\sigma
 + 8\rho^{(3)}\sum_{\sigma\neq\nu} \prod_{\eta\neq\nu,\sigma} \cos k_\eta
 + 16\rho^{(4)}\prod_{\sigma\neq\nu} \cos k_\sigma,
\ee
and
\bea 
 m_{\rm HF}(K) = &&\hm
\lambda^{(0)} 
+ 2\lambda^{(1)}\sum_\nu \cos k_\nu 
+ 4\lambda^{(2)}\sum_\nu \sum_{\sigma>\nu}\prod_{\eta\neq\nu,\sigma} 
\cos k_\eta 
\nn\\  
\label{eqmhf}
&&\hm 
+ 8\lambda^{(3)}\sum_\nu \prod_{\sigma\neq \nu} \cos k_\sigma 
+ 16\lambda^{(4)}\prod_\nu \cos k_\nu.
\eea
 The coefficients $\rho^{(a)}$ and $\lambda^{(a)}$ are determined 
by requiring that this action reproduces the 
perfect action on a finite volume with sides of length 3, and periodic 
boundary conditions, see Table \ref{table2} for two examples.

\begin{table}[t]
\begin{center}
\begin{tabular}{|c|c|c||c|c|c|}
\hline
 & massless HF & massive HF & & massless HF & massive HF \\
\hline
 $\rho^{(1)}$    & 0.136846794	& 0.054580  & $\lambda^{(0)}$ & 1.852720547  & 1.268851	 \\
 $\rho^{(2)}$    & 0.032077284	& 0.011010  & $\lambda^{(1)}$ & -0.060757866 & -0.030083	 \\
 $\rho^{(3)}$ 	 & 0.011058131	& 0.003255 & $\lambda^{(2)}$ & -0.030036032 & -0.010830	 \\
 $\rho^{(4)}$    & 0.004748991	& 0.001206 & $\lambda^{(3)}$ & -0.015967620 & -0.004716	 \\
 		 &		&	   & $\lambda^{(4)}$ & -0.008426812 & -0.002212 \\
\hline
\end{tabular}
 \caption{Coefficients $\rho^{(a)}$ and $\lambda^{(a)}$ in the hypercube 
action for massless ($m_R=0$) and massive ($m_R=1$) hypercube fermions.
}
\label{table2}
\end{center}
\end{table}

In the limit of zero momentum, the  action coefficients satisfy
\bea
\label{eqc1}
C_\nu^{\rm HF}(0) = &&\hm 2\left[ \rho^{(1)} + 6\rho^{(2)} + 12\rho^{(3)} + 
8\rho^{(4)} 
\right]  = \left(\frac{m_R}{e^{m_R}-1}\right)^2,
\\
m_{\rm HF}(0) = &&\hm 
 \lambda^{(0)} + 8\lambda^{(1)} + 24\lambda^{(2)} + 32\lambda^{(3)} + 
16\lambda^{(4)} =  \frac{m_{R}^2}{e^{m_{R}}-1},
\label{eqc2}
 \eea
where $m_{R}=E_\vecnul$ is the rest mass in HF dispersion relation.
These relations follow from the expression for the one-dimensional 
fixed point action, which can be evaluated explicitly, and ultimately 
they depend on the RG transformation used to construct the action.

The truncation involved in the construction of the hypercube action 
introduces chiral symmetry breaking and discretisation errors. If the 
truncation is justified, these effects will be small.\footnote{For studies 
of meson spectral functions using hypercube fermions in quenched QCD, 
see Refs.\ \cite{Wissel:2005pb,Bietenholz:2005rc}.}
 Following Ref.\ \cite{Bietenholz:1999km}, exact chiral symmetry can be 
restored by using the hypercube operator as the kernel for the overlap 
operator. The resulting overlap operator should inherit many of the 
properties of the kernel and, in particular, have much smaller cutoff 
effects than the standard overlap operator.

To determine the expression for the overlap hypercube propagator 
we first write the expression for the kernel, $X = D_{\rm HF} - \mu$,  
in momentum space, and 
\be
X(K) =  i\sum_\nu C_\nu^{\rm HF}(K)\gamma_\nu \sin k_\nu + b(K),
\;\;\;\;\;\;\;\;
b(K) = m_{\rm HF}(K) - \mu.
\ee
 The corresponding propagator for massive overlap fermions is obtained by 
multiplying $\sin k_\nu$ with $C_\nu^{\rm HF}(K)$ in Eq.\ (\ref{eqgen}) of 
Section \ref{secov}. Writing $k_4=iE_\kv$, the dispersion relation is 
again determined by a quadratic equation for $\cosh E_\kv$. Since the 
explicit expressions are rather lengthy, we have listed them in Appendix 
\ref{appHF}.

\begin{figure}[!p]
\centerline{\epsfig{figure=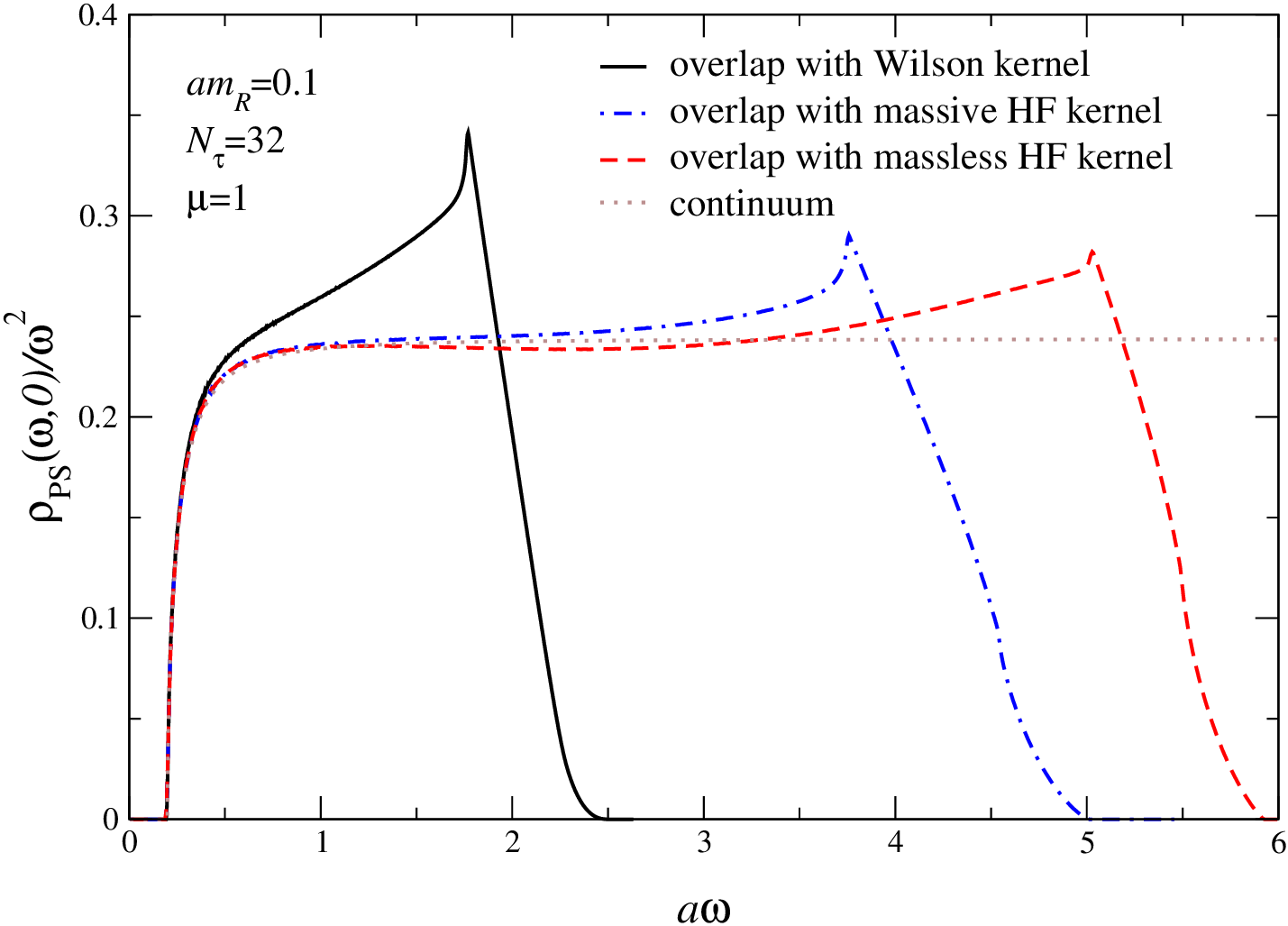,width=12cm}}
\caption{Pseudoscalar spectral functions $\rho_{\rm 
PS}(\om,\vecnul)/\om^2$ for overlap fermions with a standard and 
a massless/massive HF kernel ($\mu=1$, $m_R=0.1$, $N_\tau=32$).
}
\label{figrho_PS_HF_OV_1}
\vspace{0.3cm}
\centerline{\epsfig{figure=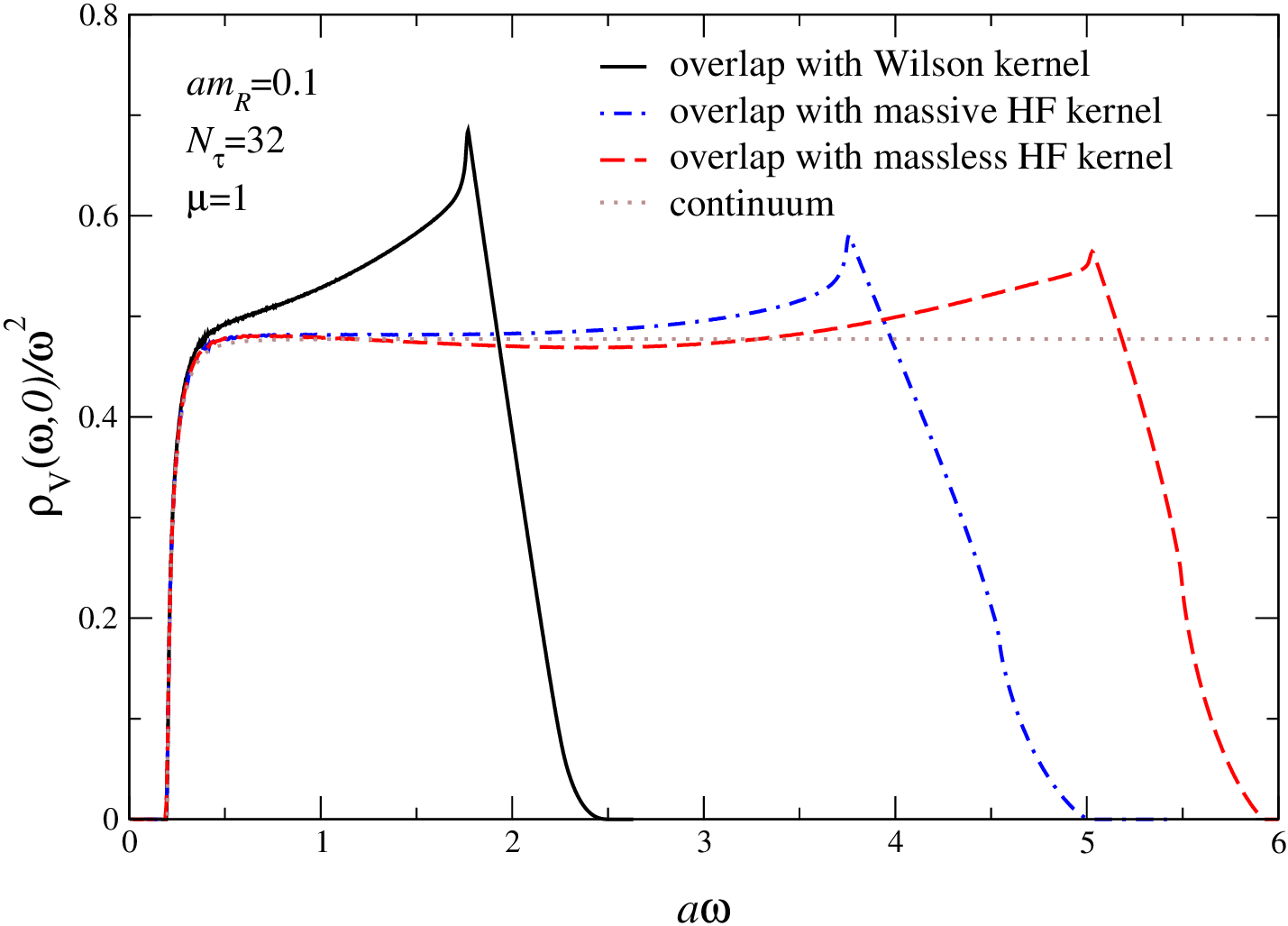,width=12cm}}
\caption{As in Fig.\ \ref{figrho_PS_HF_OV_1} for the vector spectral 
function $\rho_{\rm V}(\om,\vecnul)/\om^2$.
}
\label{figrho_V_HF_OV_1}
\end{figure}

In the limit of small $\kv$ and $m_0$, we find that the overlap HF 
dispersion relation reduces to
\be
\cosh E_\kv = 1+ \half \left( \kv^2 + m_{\rm eff}^2 \right)
+ \ldots
\ee 
where the renormalized fermion mass is
\be
\label{eqmOVHF}
m_{\rm eff} = 
\frac{ \mu-m_{\rm HF}(0) }{ \mu C_\nu^{\rm HF}(0)} m_0.
\ee
Using Eqs.\ (\ref{eqc1}, \ref{eqc2}), we find that the overlap mass $m_0$ 
receives a multiplicative renormalization for massive HF, while for 
massless HF such renormalization is absent. For 
massless overlap fermions, the dependence on the coefficients 
in the HF kernel cancels completely in the limit of small momentum. 

The resulting dispersion relations along three direction in the Brillouin 
zone are shown in Fig.\ \ref{figdisp} for overlap fermions with both a 
massless and a massive (with HF rest mass $m_R=1$) HF kernel. Note that we 
fixed the overlap rest mass at 0.1 and determined the bare mass from Eq.\ 
(\ref{eqmOVHF}). It is clear that the improved scaling of the HF kernel 
ensures agreement with the continuum dispersion relation for much larger 
momenta. The deviation at larger momenta results in the behaviour 
$\partial E_\kv/\partial\kv >1$, which is especially pronounced for the 
massless kernel.

The coefficients in the meson spectral functions are given in Eq.\ 
(\ref{eqCm}), after multiplying $C_\nu(iE_\kv,\kv)$ with $C_\nu^{\rm 
HF}(iE_\kv,\kv)$. The residue is given by Eq.\ (\ref{eqresovHF}). 
Comparison of the coefficients and the residue with their continuum 
counterparts shows again that the fermion propagator receives a 
multiplicative renormalization. Expanding the HF overlap propagator for 
small $K$ and $m_0$ yields
 \be
S(K) \approx \frac{\mu^3(\mu-m_0/2)}{\left(\mu^2 + m_0^2/4\right)^2} 
\left[\frac{\mu-m_{\rm HF}(0)}{\mu C_\nu^{\rm HF}(0)} 
\frac{-i\Kslash + m_{\rm eff} }{ K^2+m_{\rm eff}^2} 
+ \frac{1}{2\mu} \right],
\ee
 where $m_{\rm eff}$ was defined above. For overlap fermions with a 
massive HF kernel we find therefore that both the overlap mass and the 
fermion propagator are renormalized by the factor $\left[\mu-m_{\rm 
HF}(0)\right]/\mu C_\nu^{\rm HF}(0)$.

Meson spectral functions with overlap hypercube fermions are shown in 
Fig.~\ref{figrho_PS_HF_OV_1} for the pseudoscalar and Fig.\ 
\ref{figrho_V_HF_OV_1} for the vector channel. Besides a factor of 2 (see 
Table \ref{table1}), these spectral functions differ in detail around the 
knee at $a\om\sim 0.5$, which can be understood from the continuum 
expression (\ref{eqcont}). As expected, the first cusp in the spectral 
functions is shifted to substantially larger frequencies, determined by 
twice the maximal energy along the $(1,0,0)$ direction, yielding 
$a\omega\sim 3.75$ $(5.0)$ for the massive (massless) HF kernel, as can be 
seen from Fig.~\ref{figdisp} (top). As a result, the continuum behaviour 
at larger frequencies, $\rho(\om)\sim\om^2$, is better reproduced. 
However, we would like to point out that the contributions from these 
large frequencies are highly suppressed in the euclidean correlator. 
Taking for simplicity the zero temperature kernel $K(\tau,\om) = 
e^{-\omega\tau} = e^{-a\omega n_\tau}$, we find that already at the first 
time slice $K\sim e^{-5}$ when $a\om\sim 5$, demonstrating the 
insensitivity to these large frequencies.

\section{QCD with dynamical domain wall fermions}
\setcounter{equation}{0}
\label{secUKQCD}

We now build on the free field calculations and consider spectral 
functions in QCD. Most spectral function studies to date have been carried 
out in quenched QCD; for recent work, see e.g.\ Ref.\ 
\cite{Jakovac:2006sf} and references therein. A study of charmonium 
spectral functions in dynamical QCD with two flavours on highly 
anisotropic lattices can be found in Ref.\ \cite{Aarts:2006nr}. Spectral 
functions at nonzero momentum in quenched QCD are considered in Refs.\ 
\cite{Datta:2004js,Aarts:2006cq}.

 In this section, we consider spectral functions at zero momentum, 
extracted from meson correlators obtained in dynamical lattice 
simulations. These correlators are computed using the domain wall fermion 
action on 2+1 flavour background configurations employing the Iwasaki 
gauge action \cite{Iwasaki:1984cj,Iwasaki:1985we}.\footnote{Meson spectral 
functions using domain wall fermions in quenched QCD have been studied in 
Ref.\ \cite{Blum:2004zp}.} 
 This data has been generated by the RBC and UKQCD collaborations 
\cite{Allton:2006ax,Mawh,MinLi,RBC_UKQCD_8,RBC_UKQCD} on QCDOC 
\cite{Boyle:2005gf,Boyle:2003ue,Boyle:2003mj}. Here, we present results 
obtained on a $16^{3} \times 32$ lattice at $\beta=2.13$ with an inverse 
lattice spacing of $1.6~\rm{GeV}$ \cite{RBC_UKQCD}. The number of points 
in the fifth dimension is $N_s=16$.

In practical implementations of the domain wall fermion formalism it is 
important to choose a domain wall height that minimises the mixing between 
right and left-handed fermion modes at zero bare quark mass.  This mixing, 
which vanishes in the limit $N_{s} \rightarrow \infty$, induces some 
residual chiral symmetry breaking and generates, for example, an additive 
quark mass renormalisation. In free field theory the optimal value for the 
domain wall height is unity. However, in the interacting theory, the 
domain wall height receives a large additive correction and the bare 
parameter should be adjusted accordingly~\cite{Blum:1996jf, Blum:1999xi}.  
The bare domain wall height used in these simulations is $\mu=1.8$. 
Subtracting a simple mean-field estimate for the radiative corrections
\cite{Aoki:2002iq} yields a value for the domain wall height of $\mu^{\rm 
MF}=1.3029$ \cite{Boyle:2006pw}, which is much closer to unity.

In this first study we show results obtained with a light bare sea quark 
mass $m_{ud}=0.02$ and a heavier sea quark mass $m_s=0.04$. To preserve 
unitarity, the valence quarks are constrained to take the same bare mass 
values as the sea quarks.  To determine the spectral functions the Maximum 
Entropy Method \cite{Asakawa:2000tr} is applied to correlation functions 
measured on an ensemble of 70 independent gauge field configurations. The 
meson interpolating operators used are local quark field bilinears. The 
meson correlators are symmetric about the central time slice of the 
lattice. In our analysis we average the correlation functions over 
equivalent time slices and exclude the contact term at $\tau=0$. The MEM 
algorithm uses Bryan's method \cite{Bryan}. 
 As the default model we take $\rho_{\rm default}(\om,\vecnul) = 
m_0\omega^2$, where the constant $m_0$ is determined by a best fit to 
the data.

\begin{figure}[t]
\centerline{\epsfig{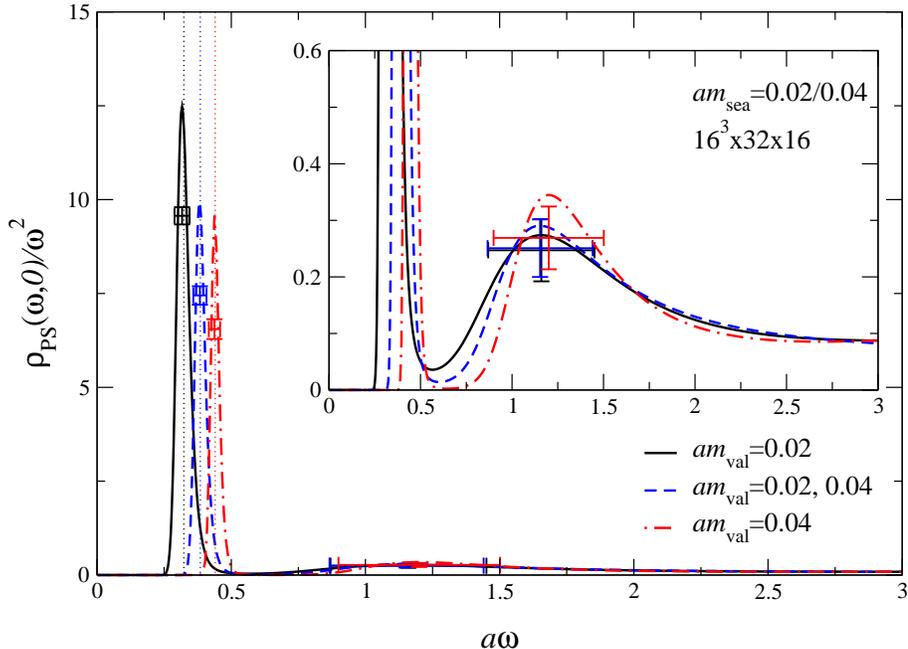}}
\caption{Pseudoscalar spectral functions in QCD with $2+1$ 
flavours of dynamical domain wall fermions, determined using the Maximum 
Entropy Method, for different values of the valence quark masses. The 
vertical dotted lines indicate groundstate masses obtained with 
conventional cosh 
fits \cite{Allton:2006ax}. The inset shows a blow-up of the second bump.
}
\label{figUKQCD_1}
\end{figure}

In Fig.\ \ref{figUKQCD_1} we show spectral functions 
obtained from pseudoscalar correlators evaluated for 
different valence quark masses. For each quark mass combination clear 
peaks are visible whose position corresponds to the energy of the lightest 
state that couples to the interpolating operators.  The horizontal 
errorbars on these peaks give a measure of the peak width, while the 
vertical errorbars are inversely proportional to the significance of the 
peak. As expected, the position of the low-lying peak increases as the 
average valence quark mass is increased. The dotted vertical lines passing 
through each of these peaks indicate the values for the ground state 
energies obtained from double cosh fits to the 
correlators~\cite{Allton:2006ax}. Therefore, for the groundstates we find 
that the results of the MEM analysis are in precise agreement with the 
results of conventional fitting techniques.

At higher frequencies, $ 1 < a\omega < 2$, second much broader bumps are 
visible. There appears to be no significant valence quark mass dependence 
of the position of these second peaks. We note that the heights of the 
bumps are of the same order as the structure due to the cusps observed in 
the free fermion calculation, for a domain wall height of 
unity.\footnote{Note, however, that we have not applied wave function 
renormalization, which affects the vertical scale. Moreover, the free 
calculation indicates a strong dependence on the domain wall height, such 
that a quantitative comparison would require knowledge of the renormalized 
domain wall height.} From our earlier analysis, we find therefore that the 
position of these bumps is not inconsistent with their identification as 
lattice artefacts. However, these peaks may also contain excited state 
contributions, but due to the width of the peaks these are difficult to 
resolve. An unambiguous way to disentangle excited state resonances from 
the contribution due to lattice artefacts is to carry out an analysis at 
different lattice spacings. With current data, this option is not yet 
available.\footnote{Unphysical structure at higher 
frequencies has also been observed in quenched simulations with Wilson 
fermions. In Refs.\ \cite{Yamazaki:2001er,Sasaki:2005ap} this was 
interpreted as bound states of Wilson doublers.}

\section{Summary}
\label{secsum}

We have analyzed meson spectral functions from lattice fermions with 
chiral symmetry. For free fermions, we have given a general prescription 
on how to construct lattice meson spectral functions from the euclidean 
fermion propagator, extending the analysis of Ref.\ \cite{Aarts:2005hg}. 
We have subsequently applied this to overlap fermions, domain wall 
fermions and overlap hypercube fermions. Lattice artefacts appear at 
(twice the) frequencies at which the lattice dispersion relation $E_\kv$ 
deviates from the continuum relation. The most pronounced effect is the 
appearance of cusps at frequencies determined by $\partial E_\kv/\partial 
\kv=0$. For most fermion formulations (Wilson, staggered, standard 
overlap, domain wall), these cusps appear at frequencies $1<a\omega<2$. In 
order to shift these artefacts to higher energies, it is necessary to use 
lattice fermions with improved scaling behaviour, such as hypercube 
fermions. In our spectral function analysis, we found that using the 
hypercube operator as a kernel in the overlap formalism indeed yields a 
formulation with good chiral and scaling behaviour, as could be 
anticipated from previous studies \cite{Bietenholz:1999km}.

From a comparison between overlap and domain wall spectral functions, we 
found that the latter have a remarkably strong dependence on the domain 
wall height, whereas the dependence on the corresponding parameter in the 
case of the overlap operator is much milder. For a domain wall height of 
unity and a small fermion mass, we found that the overlap and domain wall 
spectral functions are comparable.
                                                                                
In the final section of the paper we have performed a Maximum Entropy 
analysis of pseudoscalar spectral functions in QCD with 
dynamical domain wall fermions, using data generated by the UKQCD and RBC 
collaborations. We found good agreement between the groundstate masses, 
determined by conventional cosh fits, and the position of the peak in the 
spectral functions. At larger frequencies, $1<a\omega<2$, a second peak 
can be seen. We have argued that this structure is not inconsistent with 
the lattice artefacts found in the analytical study, although the presence 
of excited states cannot be excluded. An unambiguous way to distinguish 
(physical) excited states from (unphysical) lattice artefacts discussed 
here, is to repeat the analysis at different lattice spacings.


\vspace*{0.5cm}
\noindent
{\bf Acknowledgments.}
 We thank Chris Allton and Jonathan Clowser for providing us with a 
Maximum Entropy routine and easy access to the dynamical 
domain wall correlation functions. 
 We thank Dave Antonio, Peter Boyle, Brian Pendleton and Rob Tweedie for 
the generation of the non-perturbative DWF data, which is part of the 
UKQCD and RBC collaborations joint research programme. We also thank our 
colleagues in the UKQCD and RBC collaborations. The data was obtained 
using the QCDOC computers installed at the University of Edinburgh and 
Brookhaven National laboratory.  We thank PPARC (JIF grant 
PPA/J/S/1998/00756), RIKEN, Brookhaven National Laboratory and the U.S. 
Department of Energy for these facilities.
 G.A.\ is supported by a PPARC Advanced Fellowship. 



\appendix
\renewcommand{\theequation}{\Alph{section}.\arabic{equation}}

\section{More on overlap hypercube fermions}
\label{appHF}

In this appendix we collect some expressions for overlap hypercube 
fermions, discussed in Section \ref{secovHF}.

The pole is determined by
\be
\sum_\nu {C_\nu^{\rm HF}}^2(K)\sin^2 k_\nu = -\bar m_0^2 b^2(K), 
\;\;\;\;\;\;\;\; \;\;\;\; \;\;\;\; 
b(K)<0,
\ee
where $\bar m_0$ was defined in Eq.\ (\ref{eqmbar}).
In order to solve for the dispersion relation, we follow Ref.\ 
\cite{Karsch:2003wy} and use the notation 
\bea
\nn
C_4^{\rm HF}(K) =&&\hm \delta_\kv, \\
C_i^{\rm HF}(K)\sin k_i =&&\hm \alpha_{i\kv} + \beta_{i\kv} \cos k_4,
\nn \\
m_{\rm HF}(K) =&&\hm \kappa_{1\kv} + \kappa_{2\kv}  \cos k_4,
\eea
and
\be
\alpha^2_\kv = \sum_{i=1}^3 \alpha_{i\kv}^2,
\;\;\;\; \;\;\;\;
\beta^2_\kv = \sum_{i=1}^3 \beta_{i\kv}^2,
 \;\;\;\; \;\;\;\;
\alpha_\kv\cdot\beta_\kv = \sum_{i=1}^3 \alpha_{i\kv}\beta_{i\kv}. 
\ee
 Expressions for $\alpha_{i\kv}$, $\beta_{i\kv}$, $\delta_\kv$, and 
$\kappa_{1,2\kv}$ can easily be derived from these definitions combined 
with Eqs.\ (\ref{eqCnu}, \ref{eqmhf}). Explicit expressions are given in 
Eqs.\ (B.1-B.9) of Ref.\ \cite{Karsch:2003wy}.

Writing $k_4=iE_\kv$ yields again a quadratic equation for $\cosh E_\kv$, 
where the allowed solution is of the form
\be
\cosh E_\kv = \frac{x_\kv+\sqrt{y_\kv}}{z_\kv},
\ee
with
\bea
\nn
x_\kv = &&\hm \alpha_\kv\cdot\beta_\kv 
 + \bar m_0^2 \kappa_{2\kv}\left(\kappa_{1\kv}-\mu\right),
\\
\nn y_\kv = &&\hm \left(\alpha_\kv\cdot\beta_\kv\right)^2 + 
\left( \delta^2_\kv - \beta^2_\kv \right) 
\left( \delta^2_\kv + \alpha^2_\kv \right) \\
\nn &&\hm 
+ \bar m_0^2 \left[ 2\left(\alpha_\kv\cdot \beta_\kv \right) 
\kappa_{2\kv} \left(\kappa_{1\kv}-\mu\right)
+ \left(\kappa_{1\kv}-\mu\right)^2 
\left(\delta^2_\kv-\beta^2_\kv\right) 
- \kappa_{2\kv}^2 \left(\delta^2_\kv+\alpha^2_\kv\right) \right],
\;\;\;\;
\\
z_\kv = &&\hm \delta^2_\kv-\beta^2_\kv-\bar m_0^2 \kappa_{2\kv}^2.
\eea
The root $(x_\kv-\sqrt{y_\kv})/z_\kv < 0$ for all momenta inside the
Brillouin zone and therefore not a valid solution, in contrast to the
standard HF case.

The rest mass is given by
\be
\cosh E_\vecnul = \frac{\bar m_0^2 
\kappa_{2\vecnul}\left(\kappa_{1\vecnul}-\mu\right)+
\delta_\vecnul 
\sqrt{  
 \delta^2_\vecnul
+ \bar m_0^2 \left[ 
 \left(\kappa_{1\vecnul}-\mu\right)^2
- \kappa_{2\vecnul}^2 \right]
}}{
\delta^2_\vecnul-\bar m_0^2 \kappa_{2\vecnul}^2},
\ee
where
\bea
\kappa_{1\vecnul} = &&\hm \lambda^{(0)} + 6\lambda^{(1)} + 12\lambda^{(2)} 
+ 8\lambda^{(3)},
\\
\kappa_{2\vecnul} = &&\hm 
 2\lambda^{(1)} + 12\lambda^{(2)} + 24\lambda^{(3)} + 16\lambda^{(4)}.
\eea
The explicit expression for the residue reads
\be
\label{eqresovHF}
\frac{1}{\cE_\kv} = 
 \frac{\mu}{\mu^2+m_0^2/4}
\frac{1}{(\delta^2_\kv -\beta^2_\kv) \cosh E_\kv 
-\alpha_\kv\cdot\beta_\kv - \bar m_0^2 \kappa_{2\kv} b(iE_\kv,\kv)}
\frac{\om(iE_\kv,\kv)}{\sinh E_\kv}.
\ee
There exists no pole at $k_4=\pi-iE_\kv$, provided $b(K) = m_{\rm 
HF}(K)-\mu >0$, or
\be
\mu < \kappa_{1\kv} - \kappa_{2\kv}\cosh E_\kv.
\ee
 Since $\kappa_{1,2 \kv}$ depend on in a nontrivial manner on the 
coefficients in the HF kernel, this constraint on $\mu$ has to be verified 
on a case by case situation.


\end{document}